# CRYPTOCURRENCY PRICE PREDICTION USING TWITTER SENTIMENT ANALYSIS

Haritha G B and Sahana N B

Department of Electronics and Communication Engineering,
PES University, Bengaluru, Karnataka, India

## ABSTRACT

*The cryptocurrency ecosystem has been the centre of discussion on many social media platforms, following its noted volatility and varied opinions. Twitter is rapidly being utilised as a news source and a medium for bitcoin discussion. Our algorithm seeks to use historical prices and sentiment of tweets to forecast the price of Bitcoin. In this study, we develop an end-to-end model that can forecast the sentiment of a set of tweets (using a Bidirectional Encoder Representations from Transformers - based Neural Network Model) and forecast the price of Bitcoin (using Gated Recurrent Unit) using the predicted sentiment and other metrics like historical cryptocurrency price data, tweet volume, a user's following, and whether or not a user is verified. The sentiment prediction gave a Mean Absolute Percentage Error of 9.45%, an average of real-time data, and test data. The mean absolute percent error for the price prediction was 3.6%.*

## 1. INTRODUCTION

The frequency of cryptocurrencies-related news stories and social media posts, especially tweets, is increasing quickly along with the economic and societal impact of cryptocurrencies. Similar to traditional financial markets, there appears to be a link between media attention and the cost of cryptocurrencies. While there are many factors that influence cryptocurrency prices, it is crucial to investigate whether sentiment analysis of publicly accessible web media may help forecast whether a coin's price (or perceived value) will increase or decrease.

The concept of blockchain technology is originally described in [1]. Blockchain offers a permanent record of network transactions. A virtual currency created for payments where the sender and recipient cannot be identified is called Bitcoin (BTC), which is unrelated to governments or banks. Bitcoin's market capitalization was $381.12 billion in September 2022. The volatility of Bitcoin is ten times that of major exchange rates. [2]. There are numerous reasons for this volatility, [3] lists 22 factors that may impact volatility in the Bitcoin market. Google Trends, total circulation, consumer confidence, and the S&P 500 index are a few of these variables mentioned [3]. In this study, we investigate the feasibility of estimating the price of Bitcoin using parameters that take into account both past prices and prevailing opinions. We investigate the role the impact that these aspects have on Bitcoin's price and the various ways it can be approximated. In spite of this volatility, it is increasingly being allocated in modern investment portfolios as either an investment or as a form of currency which makes prediction of its price movements a challenging as well as imperative problem to solve.

Price prediction of cryptocurrencies using sentiment analysis is a culmination of two separate processes. Sentiment analysis can fundamentally be done using two methods: rule-based or using machine learning techniques. The rule-based method uses a lexicon of words and stated rules to classify tweets. VADER (Valence Aware Dictionary and Sentiment Reasoner) is an instance of lexicon and rule-based sentiment analysis that requires no training [4]. It produces a vector of sentiment scores: positive, negative, and neutral, whose polarities are then normalised between 0 and 1.

Machine learning techniques on the other hand analyse tweets based on algorithms like Random Forest, Support Vector Machines, etc. For sentiment prediction, [5] compares logistic regression, linear support vector machines, and Naive Bayes. For Bitcoin, they find that logistic regression performs the best: it predicted 43.9% of price increases and 61.9% of price decreases correctly.

The second segment involves price prediction of highly volatile cryptocurrencies. LSTM and GRU are RNN variations developed in a way to get around the issue of vanishing gradients and both of these algorithms have exhibited dominance in time series predictions [6]. A hybrid version of LSTM and GRU is implemented in [7] to predict crypto prices where the inputs are passed individually to both models and then through another dense activation layer to get the output.

There have been previous attempts to forecast the price of cryptocurrencies and the fluctuations of Bitcoin. [8] reported 90% accuracy and used a dataset that was scraped and the dataset was labelled using an online API. As a result, rather than forecasting price changes, their accuracy assessment was based on how well their framework aligned with the online text sentiment API. Additionally, [9] scraped tweets that were tagged "NEO" (a particular cryptocurrency) and manually labelled the tweets as positive or negative. The sentiment prediction in [9] shows a 77% accuracy and this sentiment is correlated with the price of NEO. [10] compares six BERT-based models for sentiment analysis and eight regression models for price prediction. Data is scraped from Twitter, Reddit, as well as news articles to collate a wide range of public sentiments around cryptocurrency.

There have also been attempts at using more conventional methods of making predictions using past cryptocurrency price data. [11] doubled their investment over a 60-day period by using Bayesian regression. [12] managed a bitcoin portfolio that forecasted prices using deep reinforcement learning. The value of their portfolio increased by 10 times. None of these techniques made use of news or social media data to identify trends that were not readily obvious in price history information.

Our approach involves a comprehensive dataset with unique features including the number of followers per user, volume of daily trades, etc as opposed to conventional methods of using only historical price and, if any, predicted sentiment. This implementation uses FinBERT [13] for sentiment prediction and hence evades direct usage of a rule-based approach like VADER. We, however, use VADER to label the dataset with a composite sentiment score. Using BERT provides us the additional advantage of having contextual encoding for the tweets as opposed to VADER which does not consider the context. We also propose using GRU to forecast BTC prices. This combination of two data sources and automation of both stages warrants this nomenclature of a double regression problem.

## 2. IMPLEMENTATION

The objective of our work is to leverage social media data to capture the trend around Bitcoin investments and couple that with Bitcoin prices on that given day to forecast the price movement. The project consists of three parts: (1) collecting the data streams and collating it into a single, cleaned, and comprehensive dataset, (2) passing the Tweet data through the BERT-based sentiment analysis model, and finally (3) using GRU to forecast prices using the sentiment extracted in the previous step as well as other features.

### 2.1. Dataset

The dataset used for our work consists of two data sources. One data source is historical price data of Bitcoin, extracted from Yahoo finance. This data consists of five columns: high, low, open, close, and adjusted close. The price is added on a daily basis to the dataset. "High" represents the highest price of Bitcoin for the day, and "low" represents the lowest price of Bitcoin for the day. "Open" is the price that the Bitcoin opened at and "close" is the price that it closed at for the day. "Adjusted close" represents the price of Bitcoin on that given day, adjusted against any dividends paid for the day. The second part of the dataset used is an open-sourced dataset from Kaggle [14]. It is a dataset that is daily updated and consists of real-time cryptocurrency tweets scraped. The used data is from 5 July 2021 to 4 August 2022. This part of the dataset consists of 13 features. Table 1 lists the features from the datasets, which of them were retained, and which ones we dropped due to a lack of correlation with the output.

Table 1. Feature List and Utilisation

| Sl No. | Datasets and Respective Metrics | | | |
|---|---|---|---|---|
| | Bitcoin Historical Price | | Cryptocurrency Tweets | |
| 1. | High | | user_name | ✓ |
| 2. | Low | ✓ | date | ✓ |
| 3. | Open | ✓ | text | ✓ |
| 4. | Close | ✓ | hashtags | ✓ |
| 5. | Adjusted Close | ✓ | user_followers | ✓ |
| 6. | Volume | ✗ | user_description | ✗ |
| 7. | | ✗ | user_location | ✗ |
| 8. | | | user_created | ✗ |
| 9. | | | user_friends | ✗ |
| 10. | | | user_favourites | ✗ |
| 11. | | | user_verified | ✗ |
| 12. | | | source | ✗ |
| 13. | | | is_retweet | ✗ |

The "sentiment" feature was not natively present in the dataset. Since VADER is a rule-based method of sentiment analysis, we used it to manually label our dataset. It has an F1 score of 0.96 and is reported to outperform human raters [15]. VADER takes into account negations and contractions ("not good", "wasn't good") punctuation ("good!!!"), capitalized words, text-based 'emotes' (for example: ":)"), emotion intensification (very, kind of), acronyms, and scores tweets between -1.0 (negative) and 1.0 (positive). Therefore, after merging both datasets, our dataset contained four columns for price data, five columns for Twitter data, and one column for the sentiment score. We used the prices from the time period from 5 August 2022 to 5 September 2022 to test our model

Table 2. Split of Tweets

| Training | Testing | Validation | Removed | Total |
|---|---|---|---|---|
| 1061402 | 317335 | 317335 | 206542 | 1902614 |

In the data preprocessing, tweets with irrelevant hashtags like 'giveaway' was removed. Tweets from users with less than ten followers were also removed. Tweets with zero text and only hashtags were also removed. For the sentiment analysis model, we used 1061402 training samples, 317335 testing samples, and 317335 validation samples. Table II elucidates the split of these tweets, where 'removed' is the number of tweets removed due to data augmentation and spam. The same number of tweets fro training, testing and validation were used for price prediction as well.

Our dataset had a significant degree of imbalance, according to the results of a statistical analysis. About 32% of the dataset was strictly neutral (sentiment score of 0) even though the sentiment values were continuous in nature. The 25 percentile and 50 percentile of the data were strictly zero. Out of 1.2 million tweets, 412,000+ tweets were strictly neutral, and more tweets were around this region.

To solve this, three operations were implemented as data augmentation. Half of the strictly neutral tweets were removed from the dataset. This section was undersampled to avoid overfitting. Gaussian noise was added to a quarter of the dataset to have lesser strictly neutral data. The added Gaussian noise had a mean of 0 and a standard deviation of 7% of the input signal. This was decided through trial and error and the respective MAPE values were obtained. We found that for 5% and 6%, the model was slightly overfitting towards the end of training. Thirdly, spam tweets were being classified as neutral and adding to the class imbalance. Tweets with only emojis and only hashtags were removed.

The below figures show the dataset distribution changes through augmentation.

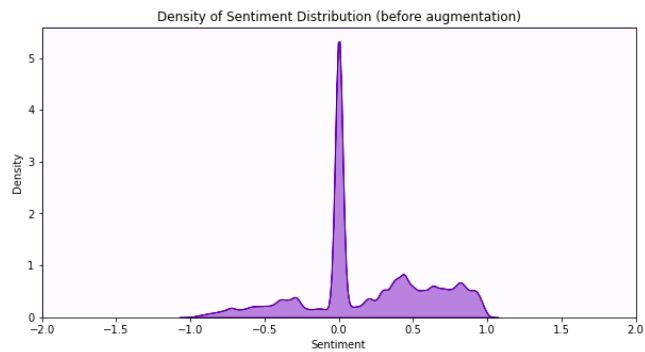

Figure 1. Dataset distribution before augmentation

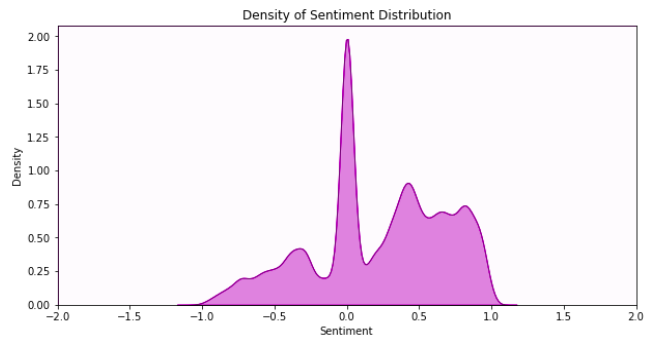

Figure 2. Dataset distribution after augmentation

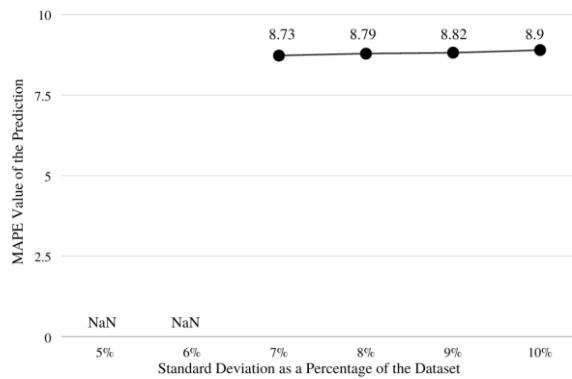

Figure 3. Comparison of MAPE Value for Different Standard Deviations of Gaussian Noise

## 2.2. Sentiment Prediction

The sentiment prediction in this paper is done using a fine-tuned FinBERT-based model. FinBert is a language model based on BERT, built to tackle NLP problems in the financial domain [13]. On various NLP tasks, pre-trained generic language models [16]–[18] have produced excellent results. Such models can be used for practically any task after being unsupervised and trained on massive quantities of text. One of the most effective language models now in use is BERT [19]. The transformer encoder is the foundation of this concept. The Transformer is a sequence-to-sequence architecture that only relies on decoder and encoder attention methods. Since the BERT design is not a sequence-to-sequence model, it ignores the decoder network and uses simply a transformer encoder (although it can be used in such tasks).

We show the fundamental BERT architecture in Figure 3. A typical word embedding vector, a position embedding vector, and a sentence embedding vector are the three representation vectors for each token that make up the token representation vector Ei, the input for the network. Since transformer models lack this understanding, the position embedding gives the model information about the token's location within the phrase. Sentiment analysis does not require a context wider than a sentence, hence the sentence vector is only employed in those situations (we consider a document as a sentence).

Each input sample in BERT is augmented with an initial artificial token called CLS. When fine-tuning the model on a text classification task, the CLS output representation will be the input to the classification layer. This is usually a softmax layer.

BERT's developers proposed two models, one smaller and one larger, with different values for the parameters L - layers, H - hidden layer size, and A - attention heads. The smaller model was called BERT BASE and had L = 12, H = 768, and E = 12, while the larger model was called BERT LARGE had L = 24, H = 1024, and E = 16. Due to our restricted computing capacity, we used the smaller BERT BASE in this study. The authors recommend fine-tuning over 4 epochs of training for fine-tuning BERT on a specific NLP task and batch sizes of 8, 16, 32, 64, and 128. They also recommend the learning rate (for Adam) to be any of: 5e-5, 3e-5, 2e-5.

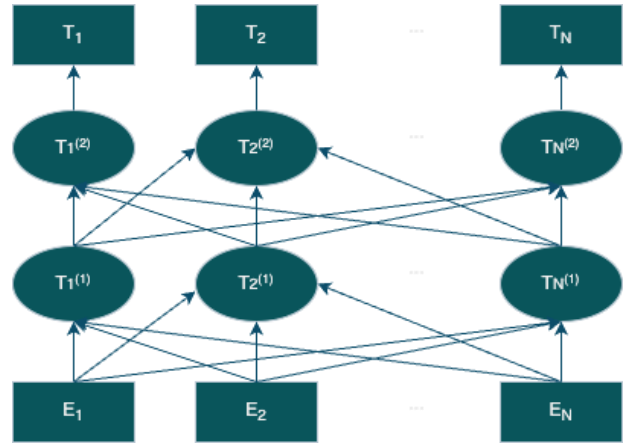

Figure 3. BERT architecture with a two-layer encoder

Through trial and error, we found that a batch size of 32, 2 epochs, and a learning rate of $5e^{-5}$ work best to avoid overfitting for our dataset.

## 2.3. Price Prediction

The basic work-flow of a Gated Recurrent Unit Network is identical to that of a basic Recurrent Neural Network; however, Gated Recurrent Unit networks differ from basic Recurrent Neural Networks in that they have gates that regulate the current input and the prior hidden state.

GRU uses what is referred to as an update gate and reset gate to address the vanishing gradient issue of a normal RNN. These are essentially two vectors that determine what data should be sent to the output. Their unique quality is that they can be taught to retain knowledge from the past without having it distorted by the passage of time or to discard information that is unrelated to the forecast.

The update gate assists the model in deciding how much historical data from earlier period steps should be transmitted to the future. The reset gate from the model is utilized to choose how much of the previous data to forget.

GRU has been used for several regression problems along with RNN and LSTMs. Both LSTMs and GRUs are similar in structure and are both used for solving the vanishing gradient problem posed by RNNs. The main distinction between GRU and LSTM is that the former has two gates—"reset" and "update," while the latter has three gates—"input," "output," and "forget." GRU has fewer gates than LSTM, making it less complicated. In our work, we have used a GRU for the forecasting of Bitcoin price from 4 August 2022 to 5 September 2022.

Computer Science & Information Technology (CS & IT) 19

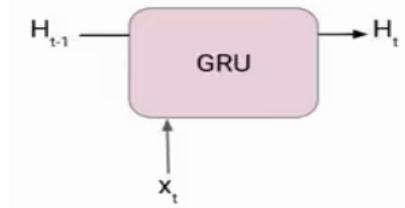

Figure 4. Structure and gates of a GRU

## 3. EXPERIMENTAL RESULTS

### 3.1. Sentiment Prediction with FinBERT

Table 2 describes the MAPE (mean absolute percentage error) value of BERT against other sentiment prediction algorithms for our dataset. Real-time testing involved scraping a stream of tweets corresponding to a particular set of daily prices outside the dataset. Test data consists of testing with tweets from the dataset scraped between 5 July 2021 to 4 August 2022. Predictions were made for two time periods: 5 August 2022 to 5 September 2022 and 5 September 2022 to 5 October 2022. MAPE Values for both were recorded and the average was found. Table 3 describes the results obtained for the price prediction of the same two time periods with comparisons using BiLSTM and GRU.

Table 3. Comparative results of Sentiment Prediction Methods

| Time Period | Mean Absolute Percentage Error | | | | | |
| --- | --- | --- | --- | --- | --- | --- |
| | Bi-LSTM | | GRU | | FinBERT-NN | |
| | *Real-time* | *Test data* | *Real time* | *Test data* | *Real time* | *Test data* |
| 5 August 2022 - 5 September 2022 | 12.32 | 11.34 | 11.47 | 9.32 | 9.91 | 8.93 |
| 5 September 2022 - 5 October 2022 | 12.01 | 11.05 | 11.57 | 9.29 | 9.78 | 9.19 |
| Average | 12.17 | 11.20 | 11.52 | 9.31 | 9.85 | 9.06 |

Table 4. GRU Price Prediction Results

| Time Period | Mean Absolute Percentage Error | |
| --- | --- | --- |
| | *Test data* | *Average* |
| 5 August 2022 - 5 September 2022 | 3.44% | 3.6% |
| 5 September 2022 - 5 October 2022 | 3.77% | |

## 3.2. Price Prediction with GRU

The below figure illustrates the prediction of the model against the real prices.

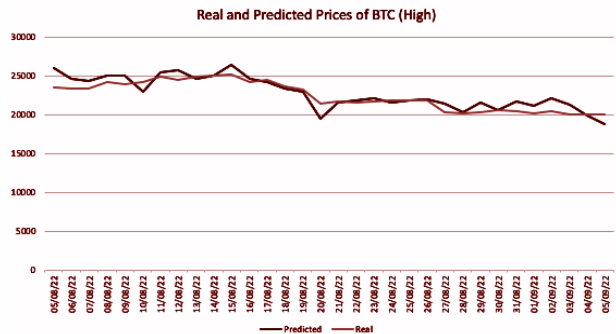

Figure 5. Comparison of real and predicted prices by model

When the model was tested with data from 5 August 2022 to 5 September 2022, the MAPE value obtained was 3.95%. Fig. 6 shows the MAPE value for each prediction that was made.

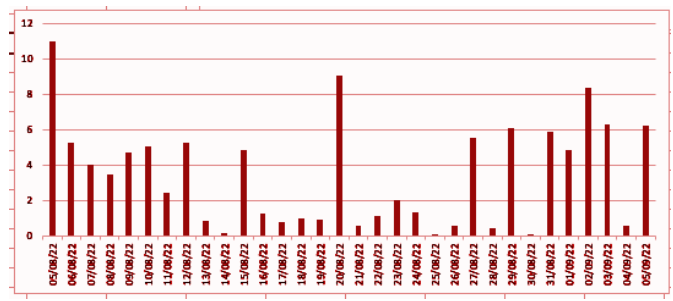

Fig. 6 MAPE Value for each prediction

The resources required to recreate this work are documented in [20].

## 4. CONCLUSIONS

This work has used the sentiment of tweets as well as corresponding daily price data to forecast the price movements of Bitcoin. The usage of FinBERT in sentiment analysis allows for contextual embedding as well as higher accuracy. The MAPE for the sentiment prediction using FinBERT for the real-time and test data averages to 9.45%  and the MAPE for the price prediction using GRU is 3.6%.

Future work will include price prediction using sentiments from multiple media: mainstream news, online articles, Reddit, and Twitter
.

**AUTHORS**

**Haritha G B**
Department of Electronics and Communication Engineering, PES University, Benagluru, Karnataka, India.

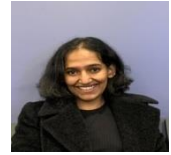

**Sahana N B**
Department of Electronics and Communication Engineering, PES University, Bengaluru , India.

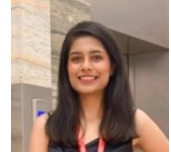